\documentclass[12pt]{article}
\usepackage{amsmath}
\usepackage{amssymb}
\usepackage{amsfonts}

\DeclareMathOperator{\sech}{sech}
\DeclareMathOperator{\real}{Re}
\newcommand{\field}[1]{\mathbb{#1}}
\newcommand{\R}{\field{R}}
\newcommand{\C}{\field{C}}

\title{
An update on $\cal PT$-symmetric complexified Scarf II potential, spectral singularities and some remarks on the rationally-extended supersymmetric partners}
\author{B Bagchi$^1$, C Quesne$^{2}$\\ 
{\small $^1$ Department of Applied Mathematics, University of Calcutta,} \\ {\small 92 Acharya Prafulla Chandra Road, Kolkata 700 009, India}\\ 
{\small $^2$ Physique Nucl\'eaireTh\'eorique et Physique Math\'ematique,  Universit\'e Libre de Bruxelles,} \\ 
{\small Campus de la Plaine CP229, Boulevard~du Triomphe, B-1050 Brussels, Belgium}\\
{\small E-mail: bbagchi123@rediffmail.com and cquesne@ulb.ac.be}}
\date{ }
\begin{document}
\baselineskip=22pt plus 1pt minus 1pt
\maketitle

\begin{abstract} 
The $\cal PT$-symmetric complexified Scarf II potential  $V(x)= - V_1 \sech^{2}x + {\rm i} V_2 \sech x \tanh x$, $V_1>0$ , $V_{2}\neq 0$ is revisited to study the interplay among its coupling parameters.  The existence of an isolated real and positive energy level that has been recently identified as a spectral singularity or zero-width resonance is here demonstrated through the behaviour of the corresponding wavefunctions and some property of the associated pseudo-norms is pointed out. We also construct four different rationally-extended supersymmetric partners to $V(x)$, which are $\cal PT$-symmetric or complex non-$\cal PT$-symmetric according to the coupling parameters range. A detailed study of one of these partners reveals that SUSY preserves the $V(x)$ spectral singularity existence.
\end{abstract}


\newpage

\section{Introduction}

After the pioneering work by Bender and Boettcher \cite{bender98} in 1998 that enforced the idea of $\cal PT$ symmetry to conjecture that the whole class of non-Hermitian Hamiltonians respecting this symmetry may exhibit (under some conditions related to $\cal PT$ being exact or spontaneously broken) real or conjugate pairs of energy eigenvalues, research in this direction has actively flourished during the past decade \cite{bender07}. Later, in an important development, Mostafazadeh \cite{mosta02} showed that the concept of $\cal PT$ symmetry is rooted in the theory of pseudo-Hermitian operators. The pseudo-Hermiticity of the Hamiltonian serves as one of the plausible necessary and sufficient conditions for the reality of the spectrum \cite{mosta08}.\par
%
%
Among the numerous models proposed on complex Hamiltonians which are $\cal PT$-symmetric/pseudo-Hermitian, the complexified Scarf II potential \cite{ahmed01a, ahmed01b, bagchi02a}
\begin{equation}
  V(x)= - V_1 \sech^{2}x + {\rm i} V_2 \sech x \tanh x, \qquad V_1 > 0, \qquad V_2 \ne 0
  \label{eq:Scarf}
\end{equation}
is particularly of interest due to a variety of reasons. First, a class of Hamiltonians having (\ref{eq:Scarf}) as its potential is not only $\cal PT$-symmetric but also $\cal P$-pseudo-Hermitian \cite{bagchi02b}. Moreover it is pseudo-supersymmetric \cite{sinha} and also nonlinearly so \cite{roy}. Second, despite being non-Hermitian in character, it is isospectral to a real potential admitting of a real discrete spectrum \cite{bagchi00a}.  Third, it is realized from the very early days of $\cal PT$ symmetry that in the framework of two non-commuting inter-connecting complex $sl(2)$ algebras \cite{bagchi02a, bagchi00b}, there are in general two series of energy levels associated with it. Note that the conventional Hermitian hyperbolic Scarf potential is rendered $\cal PT$-symmetric by complexifying one of its coupling parameters --- indeed such a complexification is responsible for the appearance of an additional series of energy levels as first pointed out by Bagchi and Quesne in \cite{bagchi00b}. The second series of bound states shows up as resonances in its Hermitian version. The $\cal PT$-symmetric Scarf II has been interpreted in terms of supersymmetry \cite{bagchi01a, levai02a} and also in the framework of an $su(1,1) \sim  so(2,1)$ algebra \cite{levai01}, as well as that of an $so(2,2)$ potential algebra \cite{levai02b}. Fourth, unlike its Hermitian counterpart the rationally-extended version of (\ref{eq:Scarf}), namely the potential
\begin{equation}
\begin{split}
  V_{\rm ext}(x) &= - (V_1 - 2a) \sech^2 x + {\rm i} (V_2 - 2b) \sech x \tanh x  \\
  & \quad - \frac{4b}{2b - {\rm i}(2a-1) \sinh x} + 2 \frac{4b^2 - (2a-1)^2}{[2b - {\rm i}(2a-1) \sinh x]^2}, 
\end{split}  \label{eq:Scarf-ext}
\end{equation}
where $a$ and $b$ are appropriately chosen real parameters known from a SUSY association with (\ref{eq:Scarf}), is free from any pole-like singularity, as already observed in a special case in \cite{bagchi09}.  Fifth, in a recent development \cite{ahmed09}, it has been shown in a general formulation that for certain conditions prevailing upon its parameters, it runs into a single zero-width resonance or a so-called spectral singularity \cite{samsonov, mosta09a, mosta09b}.\par
%
%
In this paper, we deepen our understanding on the $\cal PT$-symmetric Scarf II potential and its rationally-extended version by providing an update on them that also brings out some of their new underlying features especially the one concerning the existence of spectral singularities in the context of Ahmed's recent work \cite{ahmed09}.\par
%
%
In section 2, we start by reviewing the bound-state wavefunctions of the $\cal PT$-symmetric Scarf II potential. Such a knowledge is then used in section 3 to demonstrate the existence  of spectral singularities. The construction of rationally-extended SUSY partners is carried out in section 4. Bound-state wavefunctions and spectral singularities are presented for one of these partners in section 5. Finally, section 6 contains the conclusion.\par
%
%
\section{\boldmath Bound-state wavefunctions of $\cal PT$-symmetric Scarf II potential}

We begin by giving a direct evaluation of the bound-state wavefunctions corresponding to the potential (\ref{eq:Scarf}). Adopting the following notations:
\begin{equation}
  p = \tfrac{1}{2} \sqrt{|V_2| + V_1 + \tfrac{1}{4}}, \qquad q = \tfrac{1}{2} \sqrt{|V_2| - V_1 - \tfrac{1}{4}}, 
  \qquad s = \frac{1}{2}\sqrt{\frac{1}{4} + V_1 - |V_2|},  \label{eq:p-q-s} 
\end{equation}
we look for solutions of the type
\begin{equation}
  \psi(x) = \sech^{\lambda} x \exp[\mu \arctan(\sinh x)] \phi(y), \qquad \phi(y) \propto 
  P_{n}^{(\alpha,\beta)}(y),
\end{equation}
where $y = {\rm i} \sinh x$, $P_{n}^{(\alpha,\beta)}(y)$ are the Jacobi polynomials and $\lambda$, $\mu$, $\alpha$, $\beta$ are four constants to be determined appropriately.\par
%
%
Indeed substitution in (\ref{eq:Scarf}) yields the Jacobi form of the differential equation
\begin{equation}
  \left\{(1 - y^2) \frac{d^2}{dy^2} + [\beta - \alpha - (\alpha + \beta + 2)y] \frac{d}{dy} + n (n + \alpha +
  \beta + 1)\right\} P_n^{(\alpha, \beta)}(y) = 0  
\end{equation}
subject to the following matching conditions:
\begin{align}
  & \beta - \alpha = - 2{\rm i} \mu, \label{eq:match-1} \\
  & \alpha + \beta + 2 = 1 - 2 \lambda, \label{eq:match-2} \\ 
  & \lambda (\lambda+1) - \mu^2 = V_1, \label{eq:match-3} \\
  & (2\lambda + 1) \mu = - {\rm i} V_2, \label{eq:match-4} \\
  & \lambda^2 + E_n = - n (n + \alpha + \beta + 1). \label{eq:match-5}
\end{align}
From (\ref{eq:match-1}), (\ref{eq:match-2}) and (\ref{eq:match-5}), we get 
\begin{equation}
  \alpha= - \lambda + {\rm i} \mu - \tfrac{1}{2}, \qquad \beta = - \lambda - {\rm i} \mu - \tfrac{1}{2}, \qquad
  E_n= - (\lambda - n)^2.  \label{eq:sol-125}
\end{equation}
\par
%
%
Let us note here that since the asymptotic behaviour of $\psi(x)$ is controlled by the factor $e^{ - (\real \lambda - n)|x|}$, it will correspond to a bound state if $n < \real \lambda$, implying that in such a case $\real \lambda > 0$.\par
%
%
Turning to (\ref{eq:match-3}) and (\ref{eq:match-4}), we see that we can recast them into the combinations
\begin{equation}
  (\lambda \pm {\rm i}\mu)^{2} + (\lambda \pm {\rm i}\mu) - (V_1 \pm V_2) = 0.  \label{eq:match-34}
\end{equation}
\par
%
%
These second-degree equations have real solutions if $1 + 4(V_1 \pm V_2) \geq 0$. Hence if $|V_2| \leq V_1 +\frac{1}{4}$, both the equations yield real solutions, while if the contrary holds, i.e.\ $|V_2|> V_1 + \frac{1}{4}$, one of them furnishes real solutions and the other gives complex-conjugate roots. Interestingly, similar inequalities on $|V_2|$ emerge from the analysis of complex Lie algebras, such as sl(2), for studying the transition from real to complex eigenvalues \cite{bagchi02a}.\par
%
%
Let us study a more detailed treatment of these conditions.\par
%
%
\noindent
$\bullet \quad |V_2| \le V_1 + \tfrac{1}{4}$
\par
%
%
Solving (\ref{eq:match-34}) we get
\begin{equation}
\begin{split}
  \lambda & = - \tfrac{1}{2} + \tfrac{1}{2} \left(\epsilon_+ \sqrt{\tfrac{1}{4} + V_1 + V_2} + \epsilon_- 
        \sqrt{\tfrac{1}{4} + V_1 - V_2}\right) \in \R, \\
  \mu & = \tfrac{\rm i}{2} \left(- \epsilon_+ \sqrt{\tfrac{1}{4} + V_1 + V_2} + \epsilon_- \sqrt{\tfrac{1}{4} + 
        V_1 - V_2}\right) \in {\rm i} \R,  
\end{split}  \label{eq:sol-34}
\end{equation}
where $\epsilon_+$, $\epsilon_- = \pm{}$. As a consequence of (\ref{eq:sol-125}) and (\ref{eq:sol-34}), $E_n$ is real. It follows that if $0 < V_2 \le V_1 + \frac{1}{4}$, we have to choose $\epsilon_+ = +$, $\epsilon_- = \epsilon = \pm{}$ in order to get $\lambda > 0$. Employing the notations defined in (\ref{eq:p-q-s}), we can express
\begin{equation}
  \lambda = - \tfrac{1}{2} + p + \epsilon s, \qquad \mu = - {\rm i} (p - \epsilon s), \qquad \alpha = - 2\epsilon s,
  \qquad \beta = - 2p.
\end{equation}
\par
%
%
However, if $- V_1 - \frac{1}{4} \le V_2 < 0$ holds, we have to choose $\epsilon_- = +{}$, $\epsilon_+ = \epsilon = \pm{}$ yielding
\begin{equation}
  \lambda = - \tfrac{1}{2} + p + \epsilon s, \qquad \mu = {\rm i} (p - \epsilon s), \qquad \alpha = - 2p, \qquad 
  \beta = - 2\epsilon s.
\end{equation}
\par
%
%
On denoting the sign of $V_2$ by $\nu$, both the sub-cases given above can be considered simultaneously and we arrive at the results
\begin{equation}
\begin{split}
  & E_{n\epsilon} = - \left(p + \epsilon s - n - \tfrac{1}{2}\right)^2, \quad n = 0, 1, 2, \ldots < p + \epsilon s -
        \tfrac{1}{2} \quad \left({\rm provided\ } p + \epsilon s > \tfrac{1}{2} \right), \\
  & \lambda = - \tfrac{1}{2} + p + \epsilon s, \qquad \mu = - {\rm i} \nu (p - \epsilon s), \\
  & \alpha = - (1+\nu) \epsilon s - (1-\nu) p, \qquad \beta = - (1+\nu) p - (1-\nu) \epsilon s.      
\end{split}  \label{eq:real-E}
\end{equation}
\par
%
%
\noindent
$\bullet \quad |V_2| > V_1 + \tfrac{1}{4}$
\par
%
%
Here two cases arise. For $V_2 > V_1 + \frac{1}{4}$, we obtain
\begin{equation}
  \lambda = - \tfrac{1}{2} + \epsilon_+ p + {\rm i} \epsilon_- q \in \C, \qquad \mu = {\rm i} (- \epsilon_+ p +
  {\rm i} \epsilon_- q) \in \C. 
\end{equation}
Hence $E_n$ is now complex. Furthermore, bound states can only correspond to $\epsilon_+ = +{}$, $\epsilon_- = \epsilon = \pm{}$ and as such the parameters $\lambda$, $\mu$, $\alpha$, $\beta$ assume the form
\begin{equation}
  \lambda = - \tfrac{1}{2} + p + {\rm i} \epsilon q, \qquad \mu = - {\rm i} (p - {\rm i}\epsilon q), \qquad \alpha 
  = - 2 {\rm i} \epsilon q, \qquad \beta = - 2p.
\end{equation}
\par
%
%
On the other hand, for $V_2 < - V_1 - \frac{1}{4}$, the solutions are
\begin{equation}
  \lambda = - \tfrac{1}{2} + \epsilon_+ {\rm i} q + \epsilon_- p \in \C, \qquad \mu = {\rm i} (- \epsilon_+ {\rm i}
  q + \epsilon_- p) \in \C. 
\end{equation}
The eigenvalues $E_n$ are complex again, but the bound states now correspond to $\epsilon_- = +{}$, $\epsilon_- = \epsilon = \pm{}$. Thus it follows that
\begin{equation}
  \lambda = - \tfrac{1}{2} + p + {\rm i} \epsilon q, \qquad \mu = {\rm i} (p - {\rm i}\epsilon q), \qquad \alpha 
  = - 2p, \qquad \beta = - 2 {\rm i} \epsilon q,
\end{equation}
leading to the results
\begin{equation}
\begin{split}
  & E_{n\epsilon} = - \left(p + {\rm i} \epsilon q - n - \tfrac{1}{2}\right)^2, \qquad n = 0, 1, 2, \ldots < p - 
        \tfrac{1}{2} \quad \left({\rm provided\ } p > \tfrac{1}{2} \right), \\
  & \lambda = - \tfrac{1}{2} + p + {\rm i} \epsilon q, \qquad \mu = - {\rm i} \nu (p - {\rm i} \epsilon q), \\
  & \alpha = - (1+\nu) {\rm i} \epsilon q - (1-\nu) p, \qquad \beta = - (1+\nu) p - (1-\nu) {\rm i} \epsilon q.      
\end{split}  \label{eq:complex-E}
\end{equation}
\par
%
%
\section{\boldmath Spectral singularities of $\cal PT$-symmetric Scarf II potential}

If we set $p - \frac{1}{2} = n$ \cite{ahmed09} in equation (\ref{eq:complex-E}), then the complex energy eigenvalues $E_{n+}$ and $E_{n-}$ collapse to a single real and positive value
\begin{equation}
  E^* = q^2 = \tfrac{1}{4} \left(|V_2| - V_1 - \tfrac{1}{4}\right)  \label{eq:E*}
\end{equation}
with the conditions on $p$ translating to
\begin{equation}
  V_1 + |V_2| = 4n^2 + 4n + \tfrac{3}{4}.  \label{eq:E*-bis}
\end{equation}
\par
%
%
The corresponding wavefunctions remain solutions of the Schr\"odinger equation and are given explicitly by
\begin{equation}
  \psi_{n\epsilon}(x) = N_{n\epsilon} (\sech x)^{n + {\rm i}\epsilon q} \exp \left[- {\rm i} \nu \left(n + 
  \tfrac{1}{2} - {\rm i} \epsilon q\right) \arctan(\sinh x)\right] P_n^{(\alpha, \beta)}({\rm i} \sinh x),
\end{equation}
where
\begin{equation}
  \alpha = - (1-\nu) \left(n + \tfrac{1}{2}\right) - (1+\nu) {\rm i}\epsilon q, \qquad \beta = - (1+\nu) \left(n + 
  \tfrac{1}{2}\right) - (1-\nu) {\rm i}\epsilon q 
\end{equation}
and $N_{n\epsilon}$ are some undetermined normalization coefficients.\par
%
%
{}For $x \to \pm \infty$, they satisfy the asymptotic boundary conditions
\begin{equation}
  \psi_{n\epsilon} (x) \to N_{n\epsilon} (\pm {\rm i})^n 2^{{\rm i} \epsilon q} \exp\left[\mp {\rm i} \nu \left(n  
  + \frac{1}{2} - {\rm i} \epsilon q\right) \frac{\pi}{2}\right] e^{\mp {\rm i} \epsilon q x}.
\end{equation}
From this we infer that the solutions of the eigenvalue equation $H \psi_{q\pm}(x) = E^* \psi_{q\pm}(x)$ such that $\psi_{q\pm}(x) \to e^{\pm {\rm i} qx}$ as $x \to \pm \infty$, i.e.\ the Jost solutions, are both proportional to $\psi_{n-}(x)$, hence are linearly dependent. We therefore conclude \cite{samsonov, mosta09a, mosta09b} that $E^*$ is a spectral singularity of $H$ for the complexified Scarf II potential (\ref{eq:Scarf}).\par
%
%
Normally in $\cal PT$-spontaneously broken scenarios, complex conjugate eigenvalues develop and the energy eigenfunctions cease to be eigenstates of the $\cal PT$ operator while their pseudo-norm vanishes \cite{bagchi01b}. In the present case, we have an exceptional situation: the potential is $\cal PT$-symmetric but its wavefunctions are not so (actually ${\cal PT} \psi_{n+}(x) = \psi_{n-}(x)$); yet the corresponding eigenvalue $E^*$ is real and positive, which is a feature of a spectral singularity. Furthermore, it can be checked that after the collapse of $E_{n+}$ and $E_{n-}$, the pseudo-norm of the wavefunctions assumes a finite nonvanishing value. For $n=0$, for instance, we get
\begin{equation}
\begin{split}
  & \int_{-\infty}^{\infty} dx\, [\psi_{0\epsilon}(-x)]^* \psi_{0\epsilon}(x) = [N_{0\epsilon}|^2
         \int_{-\infty}^{\infty} dx\, \exp[- {\rm i} \nu \arctan(\sinh x)]  \\
  & = [N_{0\epsilon}|^2 \int_{-\infty}^{\infty} dx\, (\sech x - {\rm i} \tanh x) = \pi [N_{0\epsilon}|^2.  
\end{split}
\end{equation}
\par
%
%
We now make a few remarks on Ahmed's recent work \cite{ahmed09}. From the identification of the poles of the transmission amplitude for (\ref{eq:Scarf}), he was led to the following relation for the energy eigenvalues
\begin{equation}
  E_n = - \left[n + \tfrac{1}{2} - (p \pm {\rm i} q)\right]^2, \qquad n = 0, 1, 2, \ldots.
\end{equation}
We recognize $E_n$ to be also consistent with the direct determination of the same as suggested by (\ref{eq:complex-E}). Ahmed then identifies the real energy $E^*$ given by (\ref{eq:E*}) as where the spectral singularity or zero-width resonance occurs by applying Theorem 2 of \cite{mosta09a}. So both his approach and ours lead to the same conclusion, but we think that the derivation presented here looks more straightforward.\par
%
%
We would also like to point out that there is some mix-up in the assignment of the character of the parameters in \cite{ahmed09}. For instance, the notations $q$ and $s$ are mis-identified while discussing the reality of the parameters. Furthermore, the expression for the transmission amplitude is correct provided $\Gamma(1 + {\rm i} k)$ in the denominator is replaced by $\Gamma(1 - {\rm i} k)$. The confusion apparently comes from a misprint in \cite{khare} although it was clarified in \cite{levai01}.\par
%
%
\section{\boldmath Rationally-extended supersymmetric partners of $\cal PT$-symmetric Scarf II potential} 

SUSY opens the window for the construction of various types of solvable potentials \cite{cooper, junker, bagchi00c}, including the recently proposed class of completely solvable rationally-extended ones \cite{bagchi09, cq08, cq09, odake09a, odake09b, odake09c, tanaka, grandati}. Here we report on another class of rational potentials defined in (\ref{eq:Scarf-ext}), which are basically SUSY extensions of the complexified Scarf II potential (\ref{eq:Scarf}). For simplicity's sake, we henceforth restrict ourselves to $V_2 > 0$. Similar calculations could be easily carried out for $V_2 < 0$.\par
%
%
With the superpotential given by
\begin{equation}
  W(x) = a \tanh x + {\rm i} b \sech x - \frac{{\rm i} \cosh x}{{\rm i} \sinh x + c}
\end{equation}
we find that $V(x)$ and $V_{\rm ext}(x)$ are supersymmetric partners in the usual sense \cite{cooper, junker, bagchi00c}, i.e.\ $V(x) \equiv V^{(+)}(x) = W^2 - W' + E$, $V_{\rm ext}(x) \equiv V^{(-)}(x) = W^2 + W' + E$, provided the parameters $a$ and $b$ are solutions of the coupled equations
\begin{equation}
  a(a+1) + b^2 = V_1, \qquad (2a+1) b = V_2,  \label{eq:coupled}
\end{equation}
$c$ and $E$ being
\begin{equation}
  c = - \frac{2b}{2a-1}, \qquad E = - (a-1)^2.
\end{equation}
\par
%
%
The two equations in (\ref{eq:coupled}) can readily be combined to generate two corresponding ones for $a+b$ and $a-b$:
\begin{equation}
  (a+b) (a+b+1) = V_1 + V_2, \qquad (a-b) (a-b+1) = V_1 - V_2.  \label{eq:coupled-bis}
\end{equation}
The solutions of (\ref{eq:coupled-bis}) are real or complex conjugate according as the guiding discriminant $\Delta_{\pm} = 1 + 4(V_1 \pm V_2)$ is nonnegative or negative. In the case where $V_2 \le V_1 + \frac{1}{4}$, we get
$\Delta_+ > 0$, $\Delta_- \ge 0$, showing that $a+b$, $a-b \in \R$ implying $a$, $b \in \R$. In the opposite case $V_2 > V_1 + \frac{1}{4}$, we still have $\Delta_+ > 0$ but $\Delta_- < 0$ and as such $a+b \in \R$ but $a-b \in \C$. Hence $a$, $b \in \C$.\par
%
%
Let us consider the two cases in the same spirit as we did earlier for the complexified Scarf II.\par
%
%
\noindent
$\bullet \quad V_2 \le V_1 + \tfrac{1}{4}$
\par
%
%
Employing (\ref{eq:p-q-s}) we can express $a$, $b$ and $E$ as
\begin{equation}
  a = - \tfrac{1}{2} + \epsilon_+ p + \epsilon_- s, \qquad b = \epsilon_+ p - \epsilon_- s, \qquad E = - \left(
  \epsilon_+ p + \epsilon_- s - \tfrac{3}{2}\right)^2, 
\end{equation}
where $\epsilon_+$, $\epsilon_- = \pm{}$. It is obvious that according to the choice made for $\epsilon_+$ and $\epsilon_-$, we get four different $\cal PT$-symmetric extended partners.\par
%
%
The factorizing function is straightforward to derive and turns out to be
\begin{equation}
\begin{split}
  \phi(x) &\propto (\sech x)^{- \frac{1}{2} + \epsilon_+ p + \epsilon_- s} \exp[- {\rm i} (\epsilon_+ p - 
         \epsilon_- s) \arctan(\sinh x)] \\
  & \quad \times [\epsilon_+ p - \epsilon_- s - {\rm i} (\epsilon_+ p + \epsilon_- s -1) \sinh x].
\end{split}
\end{equation}
In the present case, we know that the spectrum of $V^{(+)}(x)$ is guided by two series of real eigenvalues $E^{(+)}_{n\epsilon} = E_{n\epsilon}$, given in equation (\ref{eq:real-E}), with the corresponding eigenfunctions
\begin{equation}
  \psi^{(+)}_{n\epsilon}(x) \propto (\sech x)^{- \frac{1}{2} + p + \epsilon s} \exp[- {\rm i} (p - \epsilon s)
  \arctan(\sinh x)] P_n^{(- 2 \epsilon s, - 2p)}({\rm i} \sinh x).
\end{equation}
Since $P_1^{(- 2 \epsilon s, - 2p)}({\rm i} \sinh x) = p - \epsilon s - {\rm i} (p + \epsilon s - 1) \sinh x$, we at once see that for $\epsilon_+ = +{}$ and $\epsilon_- = \epsilon$, $E = E^{(+)}_{1\epsilon}$ and $\phi = \psi^{(+)}_{1\epsilon}$. Hence the corresponding partner potential has one level less corresponding to this energy.\par
%
%
In contrast, for $\epsilon_+ = -{}$, $\phi(x)$ behaves as $\exp[(p - \epsilon_- s + \frac{3}{2})|x|] \to \infty$ at $x \to \pm \infty$, whereas $\phi^{-1}(x)$ vanishes at infinity. The partner potential has therefore  one additional level at the energy $E = - \left(p - \epsilon_- s + \frac{3}{2}\right)^2$ with a wavefunction in the form
\begin{equation}
\begin{split}
  \phi^{-1}(x) &\propto (\sech x)^{\frac{1}{2} + p - \epsilon_- s} \exp[- {\rm i} (p + \epsilon_- s) 
         \arctan(\sinh x)] \\
  & \quad \times [- p - \epsilon_- s + {\rm i} (p - \epsilon_- s +1) \sinh x]^{-1}.
\end{split}
\end{equation}
If $\epsilon_- = -{}$, this additional level lies below the two series of real eigenvalues $E^{(+)}_{n\epsilon}$. If $\epsilon_- = +{}$, it always lies below the series corresponding to $\epsilon = -{}$. However, it lies below the other series associated with $\epsilon = +{}$ only if $- \left(p - s + \frac{3}{2}\right)^2 < - \left(p + s - \frac{1}{2}\right)^2$ or $s < 1$, i.e.\ $V_2 > V_1 - \frac{15}{16}$. For $s \ge 1$, it may even coincide with $E^{(+)}_{n+}$ if $- \left(p - s + \frac{3}{2}\right)^2 = - \left(p + s - n - \frac{1}{2}\right)^2$ or $2s = n+2$, i.e.\ $\frac{1}{4} + V_1 - V_2 = (n+2)^2$. We therefore conclude that $E = E^{(+)}_{n+}$ if $V_2 = V_1 - \left(n + \frac{3}{2}\right) \left(n + \frac{5}{2}\right)$.\par
%
%
\noindent
$\bullet \quad V_2 > V_1 + \tfrac{1}{4}$
\par
%
%
All we have to do here is to replace $s$ by ${\rm i} q$. The four different partners turn out to be complex, non-$\cal PT$-symmetric potentials. A little calculation shows that in all cases, the partner potential has a finite number of pairs of complex conjugate energies and additionally a single complex energy without complex conjugate counterpart. This is not surprising for non-$\cal PT$-symmetric potentials. Finally, since
\begin{equation}
  \real E = - \left[\left(p + \tfrac{3}{2}\right)^2 - q^2\right] < \real E^{(+)}_{n\epsilon} = - \left[\left(p - n -
  \tfrac{1}{2}\right)^2 - q^2\right],
\end{equation}
in the complex plane $E$ is always well isolated at the left of the eigenvalues $E_{n\epsilon}^{(+)}$. So no double degeneracy occurs as for the real case.\par
%
%
The bound-state wavefunctions $\psi^{(-)}_{n\epsilon}(x)$ of the four different partner potentials $V^{(-)}(x) = V_{\rm ext}(x)$ can be easily found from their counterparts $\psi^{(+)}_{n\epsilon}(x)$ for the $\cal PT$-symmetric Scarf II potential $V^{(+)}(x)$ by acting with the operator $\left(E^{(+)}_{n\epsilon} - E\right)^{-1/2} \left[\frac{d}{dx} + W(x)\right]$ in the usual way \cite{cooper, junker, bagchi00c}. In the next section, we will present the results obtained for one of the four partners and prove from them the existence of spectral singularities for certain conditions prevailing on its parameters.\par
%
%
\section{Bound-state wavefunctions and spectral singularities of partners: a case study}

Let us consider more specifically the partner corresponding to the choice $\epsilon_+ = \epsilon_- = +{}$ and distinguish between the two ranges of potential parameters again.\par
%
%
\noindent
$\bullet \quad V_2 \le V_1 + \tfrac{1}{4}$
\par
%
%
The general form obtained for the partner wavefunctions reads
\begin{equation}
  \psi^{(-)}_{n\epsilon}(x) = N^{(-)}_{n\epsilon} (\sech x)^{\xi} \exp[\eta \arctan(\sinh x)] [p - s - {\rm i}(p + s
  - 1) \sinh x]^{-1} {\cal P}_{n\epsilon}({\rm i} \sinh x), \label{eq:partner-psi}
\end{equation}
where the parameters $\xi$, $\eta$ and the polynomials ${\cal P}_{n\epsilon}({\rm i} \sinh x)$ are defined below.\par
%
%
If in equation (\ref{eq:partner-psi}) $\epsilon = +{}$, then
\begin{equation}
  \xi = - \tfrac{3}{2} + p + s, \quad \eta = - {\rm i} (p-s), \quad n = 0, 2, 3, \ldots < p + s - \tfrac{1}{2}
  \quad \left({\rm provided\ } p + s > \tfrac{1}{2} \right),
\end{equation}
and ${\cal P}_{n+}(y)$ are $n$th-degree polynomials somewhat reminiscent of other ones appearing in the context of real potentials \cite{gomez04a, gomez04b}. The first two of them are given by
\begin{equation}
\begin{split}
  &{\cal P}_{0+}(y) = 1, \\
  &{\cal P}_{2+}(y) = (p+s-1) (2p+2s-3) y^2 - 2 (p-s) (2p+2s-3) y + 2 (p-s)^2 \\
  & \hphantom{{\cal P}_{2+}(y) =} - (p+s-1). 
\end{split}
\end{equation}
\par
%
%
However, if $\epsilon = -$, then
\begin{equation}
  \xi = - \tfrac{1}{2} + p - s, \quad \eta = - {\rm i} (p+s-1), \quad n = 0, 1, 2, \ldots < p - s - \tfrac{1}{2}
  \quad \left({\rm provided\ } p - s > \tfrac{1}{2} \right),
\end{equation}
and 
\begin{equation}
  {\cal P}_{n-}(y) = \hat{P}_{n+1}^{(2s-1, -2p+1)}(y)  \label{eq:exc-P}
\end{equation}
are the ($n+1$)th-degree Jacobi-type $X_1$ exceptional orthogonal polynomials of G\'omez-Ullate {\sl et al} \cite{gomez09}, which were applied to Hermitian quantum mechanics in \cite{bagchi09, cq08} and latter on generalized in \cite{cq09, odake09a}.\par
%
%
It can be easily checked that their behaviour as $x \to \pm \infty$ is given by
\begin{equation}
\begin{split}
  \psi^{(-)}_{n\epsilon}(x) \to & N^{(-)}_{n \epsilon} (\pm {\rm i})^{n + \frac{1}{2}(3 - \epsilon)} 2^{- \frac{1}{2}
        - n + p + \epsilon s} (p+s-1)^{-1} e^{\mp {\rm i}\left[p - \frac{1}{2} - \epsilon \left(s - \frac{1}{2}\right)
        \right] \frac{\pi}{2}} \\
  & \times e^{\pm \left(n + \frac{1}{2} - p - \epsilon s\right) x},
\end{split} \label{eq:boundary}
\end{equation}
so that they describe bound states. Their pseudo-norm $N^{(-)}_{n\epsilon}$ could be expressed in terms of that for the $\cal PT$-symmetric Scarf II wavefunctions, $N^{(+)}_{n\epsilon}$.\par
%
%
\noindent
$\bullet \quad V_2 > V_1 + \tfrac{1}{4}$
\par
%
%
The wavefunctions are given by equations (\ref{eq:partner-psi})--(\ref{eq:exc-P}) where we substitute ${\rm i} q$ for $s$. The same replacement in equation (\ref{eq:boundary}) shows that they remain bound-state wavefunctions. However, since the potential now breaks $\cal PT$ symmetry, the concept of pseudo-norm looses its signicance.\par
%
%
In addition, we observe that if in the corresponding spectrum
\begin{equation}
\begin{split}
  E^{(-)}_{n+} & = - \left(p + {\rm i} q - n - \tfrac{1}{2}\right)^2, \qquad n = 0, 2, 3, \ldots < p - 
        \tfrac{1}{2} \quad \left({\rm provided\ } p > \tfrac{1}{2} \right), \\
  E^{(-)}_{n-} & = - \left(p - {\rm i} q - n - \tfrac{1}{2}\right)^2, \qquad n = 0, 1, 2, \ldots < p - 
        \tfrac{1}{2} \quad \left({\rm provided\ } p > \tfrac{1}{2} \right), 
\end{split}
\end{equation}
we set $p - \frac{1}{2} = n$, then for $n=0$, 2, 3,~\ldots, the two complex energy eigenvalues $E^{(-)}_{n+}$ and $E^{(-)}_{n-}$ collapse to the real and positive value $E^*$ given in (\ref{eq:E*}), while for $n=1$ there is a single complex eigenvalue $E^{(-)}_{1-}$ that becomes equal to the same. The associated conditions (\ref{eq:E*-bis}) on $V_1$ and $V_2$ yield the relations
\begin{equation}
  V'_1 + V'_2 = 4 n^2 - \tfrac{1}{4}
\end{equation}
connecting the coefficients  $V'_1 = V_1 - 2a$, $V'_2 = V_2 - 2b$ of the first two terms on the right-hand side of (\ref{eq:Scarf-ext}).\par
%
%
{}Furthermore, it is straightforward to see that the wavefunction $\psi^{(-)}_{n-}(x)$, which exists for any $n=0$, 1, 2,~\ldots, satisfies the asymptotic boundary conditions
\begin{equation}
  \psi^{(-)}_{n-}(x) \to N^{(-)}_{n-} (\pm{\rm i})^{n+2} 2^{- {\rm i}q} \left(n - \tfrac{1}{2} + {\rm i}q\right)
  ^{-1} e^{\mp {\rm i}\left(n - \frac{1}{2} + {\rm i}q\right) \frac{\pi}{2}} e^{\pm {\rm i}qx} 
\end{equation}
as $x \to \pm \infty$, hence is proportional to both Jost solutions. This again identifies $E^*$ as a spectral singularity. We have therefore shown that for the considered partner, SUSY preserves the existence of spectral singularities.\par
%
%
\section{Conclusion}

To summarize, we have derived from the first principles the bound-state wavefunctions of the complexified Scarf II potential and the associated energy levels. For a specific condition on its underlying parameters holding, our results reveal the existence of an isolated real and positive energy level that also coincides with the determination of the same from a recent study of the poles of the transmission amplitude. However we identify it here as a spectral singularity directly from the behaviour of the corresponding wavefunctions and we point out some interesting property of the associated pseudo-norms.\par
%
%
We have also constructed new completely solvable rationally-extended partners to the complexified Scarf II and we have solved for their spectrum. Depending upon the prevailing conditions on the coupling parameters, our results point to either four different $\cal PT$-symmetric partners or complex non-$\cal PT$ ones. The behaviour of the associated energy levels is shown to be much more complicated than those for the real potentials considered in \cite{bagchi09, cq08}. As a case study we have explicitly constructed the wavefunctions for one of the four partners and demonstrated the existence of a spectral singularity for the same real and positive energy as that characterizing the complexified Scarf II potential, thereby showing the preservation of this property under SUSY.\par
%
%
\section*{Acknowledgment}

One of us (BB) thanks Prof. Kalipada Das for some valuable comments.\par
%
%

\end{document}